\documentclass[article,twocolumn,superscriptaddress,aps,nofootinbib]{revtex4-1}

\usepackage[normalem]{ulem}
\usepackage[normalem]{ulem} 
\usepackage{subfigure} 
\usepackage{graphicx}
\usepackage{dcolumn}
\usepackage{bm}
\usepackage{amsmath,amsfonts,amssymb,slashed,upgreek,color}
\usepackage{cancel}
\usepackage{ulem}
\usepackage[citecolor=blue]{hyperref}
\usepackage{comment}
\usepackage{graphics}
\usepackage{hyperref}
\usepackage[dvipsnames]{xcolor}
\usepackage{adjustbox}
\usepackage{float}
\usepackage{multirow}
\usepackage{booktabs}

\def\be{\begin{equation}}
\def\ee{\end{equation}}
\def\ba{\begin{eqnarray}}
\def\ea{\end{eqnarray}}

\begin{document}

\title{Extending MGCAMB tests of gravity to nonlinear scales}

\author{Zhuangfei Wang}
\affiliation{Department of Physics, Simon Fraser University, Burnaby, BC, V5A 1S6, Canada}

\author{Daniela Saadeh}
\affiliation{Institute of Cosmology $\&$ Gravitation, University of Portsmouth, Portsmouth, PO1 3FX, UK}

\author{Kazuya Koyama} 
\affiliation{Institute of Cosmology $\&$ Gravitation, University of Portsmouth, Portsmouth, PO1 3FX, UK}

\author{Levon Pogosian} 
\affiliation{Department of Physics, Simon Fraser University, Burnaby, BC, V5A 1S6, Canada}

\author{Benjamin Bose}
\affiliation{Institute for Astronomy, University of Edinburgh, Royal Observatory, Blackford Hill, Edinburgh, EH9 3HJ, U.K.}

\author{Lanyang Yi}
\affiliation{University of Chinese Academy of Sciences, Beijing, 100049, P.R.China}
\affiliation{National Astronomical Observatories, Chinese Academy of Sciences, Beijing, 100101, P.R.China}

\author{Gong-Bo Zhao}
\affiliation{National Astronomical Observatories, Chinese Academy of Sciences, Beijing, 100101, P.R.China}
\affiliation{University of Chinese Academy of Sciences, Beijing, 100049, P.R.China}
\affiliation{Institute for Frontiers in Astronomy and Astrophysics, Beijing Normal University, Beijing, 102206, P.R.China}

\begin{abstract}

Modified Growth with {\tt CAMB} ({\tt MGCAMB}) is a patch for the Einstein-Boltzmann solver {\tt CAMB} for cosmological tests of gravity. Until now, {\tt MGCAMB} was limited to scales well-described by linear perturbation theory. In this work, we extend the framework with a phenomenological model that can capture nonlinear corrections in a broad range of modified gravity theories. The extension employs the publicly available halo model reaction code {\tt ReACT}, developed for modeling the nonlinear corrections to cosmological observables in extensions of the $\Lambda$CDM model. The nonlinear extension makes it possible to use a wider range of data from large scale structure surveys, without applying a linear scale cut. We demonstrate that, with the 3$\times$2pt Dark Energy Survey data, we achieve a stronger constraint on the linear phenomenological functions $\mu$ and $\Sigma$, after marginalzing over the additional nonlinear parameter $p_1$, compared to the case without the nonlinear extension and using a linear cut. The new version of {\tt MGCAMB} is now forked with {\tt CAMB} on GitHub allowing for compatibility with future upgrades.
\end{abstract}

\maketitle

\section{Introduction}

The discovery of the accelerated expansion of the Universe~\cite{Perlmutter:1998np, Riess:1998cb}, the unknown nature of dark matter and the apparent fine-tuning needed to reconcile the observed value of the cosmological constant, $\Lambda$, against predictions from quantum field theory~\cite{Weinberg:1988cp,Burgess:2013ara} all stimulated exploration of alternative theories of gravity. The interest in extensions of General Relativity (GR) is further fuelled by the opportunities to test gravity at cosmological scales offered by the existing and forthcoming cosmological surveys. 

Modified Growth with {\tt CAMB} ({\tt MGCAMB}) is a patch for the widely used Einstein-Boltzmann solver {\tt CAMB}~\cite{Lewis:1999bs,camb} that enables tests of modified gravity (MG) and dark energy (DE) models. It was first introduced in 2008~\cite{Zhao:2008bn} and upgraded in 2011~\cite{Hojjati:2011ix}, 2019~\cite{Zucca:2019xhg} and 2023~\cite{Wang:2023tjj}. {\tt MGCAMB} computes predictions for the standard cosmological observables based on a phenomenological framework that allows for modified relations between the gravitational potentials and the matter density perturbation, and, like {\tt CAMB}, can be used with Monte-Carlo Markov Chain (MCMC) samplers {\tt CosmoMC}~\cite{Lewis:2002ah,cosmomc} and {\tt Cobaya}~\cite{Lewis:2002ah,Torrado:2020dgo,cobaya}. {\tt MGCAMB} can be used to explore a wide range of scalar-tensor theories of gravity under the quasi-static approximation (QSA), both in terms of generic parameterizations~~\cite{Zhao:2008bn} and specific models such as the Hu-Sawicki~\cite{Hu:2007nk} $f(R)$, the symmetron~\cite{Hinterbichler:2010es} and the dilaton~\cite{Damour:1990tw,Damour:1994zq,Brax:2012gr} models, as well as the purely phenomenological parameterizations of departures from GR introduced by the Planck ~\cite{Ade:2015rim} and Dark Energy Survey (DES)~\cite{DES:2018ufa} collaborations. 

The {\tt MGCAMB} framework is based on linear perturbation theory, with the following modifications of the perturbed Einstein equations relating the Newtonian gauge gravitational potentials $\Psi$ and $\Phi$ to the perturbations in the matter stress-energy: 
\ba
k^2 \Psi=-4 \pi G \ \mu(a, k) \ a^2 [\rho \Delta+3(\rho+P) \sigma ]
\label{eq:Poisson1} \, ,  \\
k^{2}(\Phi+\Psi)=-4 \pi G \ \Sigma(a,k) \ a^{2} [2\rho \Delta+3(\rho+P) \sigma] \, , 
\label{eq:Poisson2}
\ea
where $\rho \Delta \equiv \sum_i \rho_i \Delta_i$ and $(\rho+P) \sigma \equiv \sum_i (\rho_i+P_i) \sigma_i$, $\Delta_i$, $\sigma_i$, $\rho_i$ and $P_i$ are the comoving matter density contrast, the anisotropic stress, background density and pressure with $i \in \{b,c,\gamma,\nu\}$ standing for baryons, cold dark matter, photons and neutrinos, respectively. The phenomenological functions $\mu\equiv\mu(k,a)$ and $\Sigma\equiv\Sigma(k,a)$ in Eqs.~\ref{eq:Poisson1}-\ref{eq:Poisson2} regulate the response of non-relativistic and relativistic matter to gravity, respectively: in fact, in this formalism, photons experience an effective gravitational constant $G_{\mathrm{eff,rel}}(k,a) = G\times \Sigma(k,a)$, whereas matter is subject to $G_{\mathrm{eff,nonrel}}(k,a) = G \times \mu(k,a)$, where $G$ is Newton's constant. Similarly to parameterizations of the equation of state, which encode deviations of the expansion history from a pure $\Lambda$CDM model, $\mu$ and $\Sigma$ characterise linear deviations from standard growth. They are particularly useful in that $\mu\neq\Sigma$ is a key prediction of several modified gravity models \cite{Pogosian:2016pwr}. To make the comparisons with previously published constraints simple, we will focus on the DES parameterization~\cite{DES:2018ufa}:
\ba
\mu=1+\mu_{0} \frac{\Omega_{\mathrm{DE}}(a)}{\Omega_{\mathrm{DE}, 0}} \\
 \Sigma=1+\Sigma_{0} \frac{\Omega_{\mathrm{DE}}(a)}{\Omega_{\mathrm{DE}, 0}},
\ea
where $\mu_{0}$ and $\Sigma_{0}$ are the present values of $\mu$ and $\Sigma$, respectively, and $\Omega_{\mathrm{DE}}(a)=\rho_{\mathrm{DE}} / \rho_{\mathrm{tot}}$ is the dark energy density. 


In this paper, we follow the conventions of \cite{Ma:1995ey} and express the perturbed Friedman-Lemaître-Robertson-Walker (FLRW) metric in the Newtonian gauge:
\be
d s^{2}=a^{2}(\tau)\left[-(1+2 \Psi) d \tau^{2}+(1-2 \Phi) d x^{i} d x_{i}\right] \ .
\label{eq:separation}
\ee

The $\mu-\Sigma$ parameterization was extensively used in tests of gravity on cosmological scales (see e.g. Refs.~\cite{Pogosian:2021mcs,Andrade_2024}). Albeit powerful, this formalism can only be used within the range of validity of linear perturbation theory. On the other hand, a significant portion of the information contained in the data from large scale structure (LSS) surveys is in the correlations over scales $\lesssim 10$ Mpc, where growth is highly nonlinear~\citep{LSST:2008ijt,Euclid:2019clj,Euclid:2023rjj,SpurioMancini:2023mpt}. While in the $\Lambda$ Cold Dark Matter ($\Lambda$CDM) model the nonlinear corrections can be accurately predicted using the Halofit model~\cite{Smith:2002dz,Takahashi:2012em} (with the latest version of {\tt CAMB} adopting HMcode 2020~\cite{Mead:2020vgs}), this approach cannot be used for MG models. Because of this limitation, it was previously necessary to apply cuts to the data where they could not be reliably modelled by linear theory~\cite{Zucca:2019xhg}. In this study, we overcome this limitation by equipping {\tt MGCAMB} with the capability to compute observables at the nonlinear scales.


Refs.~\cite{Cataneo:2018cic,Bose:2020wch,Bose:2022vwi} introduced an approach, based on the halo model, to characterise the effects of nonlinearities in modified gravity theories: the \emph{halo model reaction} (HMR). In the same work, that formalism was also used to test the DGP and $f(R)$ theories at percent-level accuracy. The HMR approach was also compared to $N$-body simulations in phenomenological extensions of GR in \cite{Srinivasan:2021gib,Srinivasan:2023qsu}, showing good agreement. Recently, the HMR was applied to the forecasts of cosmic shear for Stage-IV surveys,~\cite{Tsedrik:2024cdi} which adopted the linear parametrization of MG proposed in \cite{Peebles:1980,Linder:2005in,Linder:2007hg}.

However, until now, the HMR approach has not been applied to cosmological tests of gravity using the $\mu$-$\Sigma$ parametrization used, for example, in the DES analysis \cite{DES:2018ufa}. In this work, we extend {\tt MGCAMB} to work with the HMR at nonlinear scales in this parameterization. As we will show, this addition results in stronger constraints on $\mu$ and $\Sigma$, even after marginalizing over the additional HMR parameters. Although we focus on the DES parametrization of $\mu$ and $\Sigma$ in this paper, the nonlinear extension works for all the models of $\mu$ and $\Sigma$ implemented in {\tt MGCAMB}. 

In what follows, we introduce the HMR method in Sec.~\ref{sec:react}. Sec.~\ref{sec:model} describes the implementation of the HMR in {\tt MGCAMB}, whereas we present the cosmological likelihoods used in our demonstration in Sec.~\ref{sec:demo}. We conclude with a summary in Sec.~\ref{sec:summary}.

\section{Modelling the effect of nonlinearities on the matter power spectrum} 

\label{sec:react}

In this section, we introduce the HMR method and the code used to implement it, {\tt ReACT}\footnote{\url{https://github.com/nebblu/ACTio-ReACTio}}. We refer the interested reader to Refs.~\cite{Cataneo:2018cic,Bose:2020wch,Bose:2022vwi} for further details.

\subsection{The halo model reaction method} 

To extract the information contained in the nonlinear scales, and use it to constrain models beyond $\Lambda$CDM, we need to modify the standard halo model. We can introduce a \emph{reaction function} $\mathcal{R}(k, z)$, defined as~\cite{Cataneo:2018cic}:
\be 
\mathcal{R}(k, z) \equiv \frac{P_{\text {NL}}(k, z)}{P^{\mathrm{pseudo}}(k, z)}\label{eq:reaction1} \, .
\ee
 Here, $P_{\text {NL}}(k, z)$ is the nonlinear power spectrum for the target cosmology, whereas the denominator contains the `pseudo' spectrum, which is defined as a nonlinear $\Lambda$CDM power spectrum whose linear clustering matches the modified cosmology at the target redshift $z$, i.e.:
\be
P_{\mathrm{L}}^{\mathrm{pseudo}}(k, z)=P_{\mathrm{L}}^{\mathrm{MG}}(k, z)\, .
\ee
We can produce a prediction for Eq.~\eqref{eq:reaction1} by assuming the halo model to construct the spectra. In this case the spectra are given by a sum of the 2-halo, or linear, term in the power spectrum $P_{\mathrm{L}}(k, z)$, and the 1-halo term, $P_{1h}(k, z)$ and $P_{1h}^{\mathrm{pseudo}}(k, z)$ for MG and pseudo cosmologies respectively. The reaction function represents all the corrections to the pseudo power spectrum coming from nonlinear beyond-$\Lambda$CDM physics. The choice of modelling MG corrections to a pseudo cosmology, rather than a $\Lambda$CDM one, was outlined in Ref.~\cite{Cataneo:2018cic}, which showed that the mass functions in the MG and pseudo cosmologies become quite similar, providing a smoother transition between inter- and intra-halo regimes. We direct the reader to Refs.~\cite{Cataneo:2019fjp,Bose:2020wch,Bose:2022vwi} for the construction of the reaction function using halo model quantities. Note that in the halo model reaction, 1-loop standard perturbation theory (SPT) is also employed to calibrate the 2-halo term to the 1-loop prediction, improving the overall prediction in the quasi-nonlinear regime. In this work we do not consider this calibration: Ref.~\cite{Bose:2022vwi} showed that its effects are negligible for modifications to gravity inducing a scale-independent shift in the linear growth factor, as considered here.

\subsection{The nonlinear phenomenological parameterization} 
\label{sec:parameterization}

Since we need to compute the matter power spectrum, a natural starting point is the Poisson equation, describing the relation between gravitational metric potentials and matter density contrast. In linear perturbation theory, which is the default framework of {\tt MGCAMB}, we have the modified equations given by Eqs.~(\ref{eq:Poisson1}) and (\ref{eq:Poisson2}). These can be extended to describe perturbations on smaller cosmological scales, which can be generally separated into quasi-nonlinear scales and fully nonlinear scales as ~\cite{Bose:2022vwi}:
\begin{flalign}
    & k^2 \Psi_{\mathrm{QNL}} = -4 \pi G \mu(a, k) a^2 \rho_m \delta_{\mathrm{QNL}}(k, a) + S(k, a) & \,, \\
    & k^2 \Psi_{\mathrm{NL}}(k, a) = -4 \pi G [1+\mathcal{F}(k, a)] a^2 \rho_m \delta_{\mathrm{NL}}(k, a), &
\end{flalign}
where the labels QNL and NL denote `quasi-nonlinear' and `nonlinear', respectively. Here, $S(k, a)$ is a source term containing the 2nd-order and 3rd-order perturbative contributions characterizing the modifications to the Poisson equation on quasi-nonlinear scales. Instead, the function $\mathcal{F}(k, a)$ in the second equation captures the fully nonlinear modification to the Poisson equation. 

In~\cite{Bose:2022vwi}, two different approaches were proposed to derive quantitative prediction for $\delta_{\mathrm{NL}}$, and the relation between the function $\mu$ and the nonlinear modification $\mathcal{F}(k, a)$: the parameterized post-Friedmannian framework~\cite{Lombriser:2016zfz} and a simpler phenomenological parameterization that we will adopt below. The latter approach has the benefit of being simple, having only a few free parameters, while still being able to reproduce the nonlinear effects in representative modified gravity theories.

In the phenomenological parameterization, the function $\mathcal{F}(k, a)$ is taken to be the error function (Erf), which was shown to reproduce well the general profile of the effective gravitational constant in a variety of modified gravity theories~\cite{Bose:2022vwi}. It also allows for a smooth transition from the unscreened to the screened regime. The specific form is taken to be
\be \label{eq:erf1}
\mathcal{F}_{\text {Erf }}=\operatorname{Erf}\left[a y_{\mathrm{h}} 10^{\bar{J}}\right] \times(1-\mu(\hat{k}, a)),
\ee 
where
\begin{eqnarray} 
\hat{k}&=&\frac{10^{p_{4}}}{a^{2} y_{\mathrm{h}} R_{\mathrm{th}}} \ , \\  
\bar{J}&=&p_{1}-p_{2} \log \left(R_{\mathrm{th}}\right)+p_{3} \log \left(a y_{\mathrm{env}}\right) \, , \label{eq:erf23}
\end{eqnarray}
and where $y_{\rm h} = ( R_{\rm th} /a )/ (R_{\rm i} /a_{\rm i})$, $R_{\rm th}$ being the comoving halo top-hat radius and the subscript `i' indicating the initial time. In Eq.~\eqref{eq:erf23}, $y_{\rm env}$ is the normalised radius of the environment. Note that, if $\mu \sim 1$, as in the GR limit, then $1+\mathcal{F}_{\text {Erf }} \sim 1$, and the nonlinear correction to the modified perturbations also disappears.

There are four additional parameters in Eqs.~\ref{eq:erf1}-\ref{eq:erf23}, characterising the nonlinear regime, $p_1$ - $p_4$, representing MG phenomenological effects. The first one, $p_1$, parameterizes the strength of screening, $p_2$ and $p_3$ quantify the mass and environmental dependencies, respectively, whereas $p_4$ sets the scale of Yukawa suppression~\cite{Yukawa:1935xg}. The QNL correction, described by $S(k,a)$, was shown in \cite{Bose:2022vwi} to have a negligible effect for scale-independent MG theories, while it can have up to a 2\% effect on the power spectrum for the Hu-Sawicki $f(R)$ model. We will ignore the QNL correction in the current implementation.

\section{The nonlinear extension and other upgrades of MGCAMB} 
\label{sec:model}

In this Section, we describe how we interfaced {\tt ReACT}~\cite{react}, a publicly available code for computing the reaction function, within {\tt MGCAMB}. We also describe the other upgrades implemented in the latest version of {\tt MGCAMB} and {\tt MGCobaya}~\cite{mgcobaya,Wang:2023tjj}.

\subsection{MGCAMB with ReACT} \label{Sec:MGCAMB-with-ReACT}

Since both {\tt ReACT} and {\tt MGCAMB} come with Python wrappers, making them work together amounts to establishing an appropriate Python interface. As mentioned in Sec.~\ref{sec:parameterization}, the nonlinear modification function $\mathcal{F}(k, a)$ depends on the linear modification $\mu(k,a)$, which is natively computed within {\tt MGCAMB} for several models. Therefore, the same models need to be implemented in the functions that compute $\mu$ within {\tt ReACT}.

To achieve this, we add a new ``case'', labeled 14, to the 13 cases available within {\tt ReACT}~\cite{react}. Then, we use a wrapper function to pass the parameters of the $\mu(k,a)$ functions from {\tt MGCAMB} to {\tt ReACT}. Since {\tt ReACT} can also output the modified linear power spectrum, we check that the linear predictions of {\tt MGCAMB} and {\tt ReACT} agree, so as to validate our implementation. As a representative case, we compare the linear power spectra for the DES parameterization with $\mu_0=0.4,\Sigma_0=0.1$, finding $0.1\%$ agreement between the two codes.


We also implement the dynamical dark energy options, including the cubic-spline parameterization of the dark energy density~\cite{Pogosian:2021mcs,Raveri:2021dbu,Wang:2023tjj}, to make it possible to use {\tt ReACT} with the full range of models implemented in {\tt MGCAMB}.

In {\tt MGCAMB}, we use the function call {\tt compute\_reaction\_nu\_ext}, which is an option in the Python interface of {\tt ReACT}, to compute the reaction function with the contribution of massive neutrinos. As input, this function requires three linear power spectra: $P_{\mathrm{MG}}^{\mathrm{total}}(k)$, $P_{\mathrm{MG}}^{\mathrm{cb}}(k)$ and $P_{\mathrm{\Lambda CDM}}^{\mathrm{cb}}(k)$, under the assumption of a $\Lambda$CDM background. They can be easily computed by {\tt MGCAMB}. Separately, we add a new function, named {\tt get\_react\_function}, into the {\tt MGCAMB} nonlinear Python module, to return the result of the reaction function and get {\tt MGCAMB} to compute the pseudo power spectrum ${P_{\mathrm{HM}}^{\mathrm{pseudo}}(k, z)}$ in Eq.~\eqref{eq:reaction1}. We then adopt the commonly-used interpolator function in {\tt CAMB} and {\tt MGCAMB} to get the fully nonlinear power spectrum $P_{\text {NL}}(k, z)$. 

{\tt ReACT} only provides nonlinear corrections for the matter power spectrum, not the Weyl potential $W = (\Phi+\Psi)/2$. Thus, $P_{WW}(k)$ and $P_{W \delta}(k)$, needed for interpreting data from weak lensing surveys,  must be calculated separately. We assume that the relation between $W$ and $\delta$ is the same as the one on linear scales and is given by Eq.~(\ref{eq:Poisson2}). With this assumption, we can compute $P_{WW}(k)$ and $P_{W \delta}(k)$ from the matter power spectrum $P_{\delta \delta}(k)$. 

In addition to the parameters of the functions $\mu$ and $\Sigma$, we have four parameters, $p_1$ - $p_4$, specifying the nonlinear correction. Although we interface all of them within {\tt MGCAMB}, in what follows, we take $p_1$ as the only parameter for the nonlinear regime, for simplicity, setting $p_2=p_3=p_4=0$ as in Ref.~\cite{Tsedrik:2024cdi}. This is justified when using the DES parameterization (where $\mu$ only depends on time), since for scale-independent MG models the screening has no environmental or halo mass dependence, meaning it also has no dependence on $p_2$ or $p_3$. In models with no scale dependence, such as those the DES parameterization belongs to, there is also no Yukawa suppression, and hence no dependence on $p_4$.

\begin{figure}
    \centering
    \includegraphics[width = .5\textwidth]{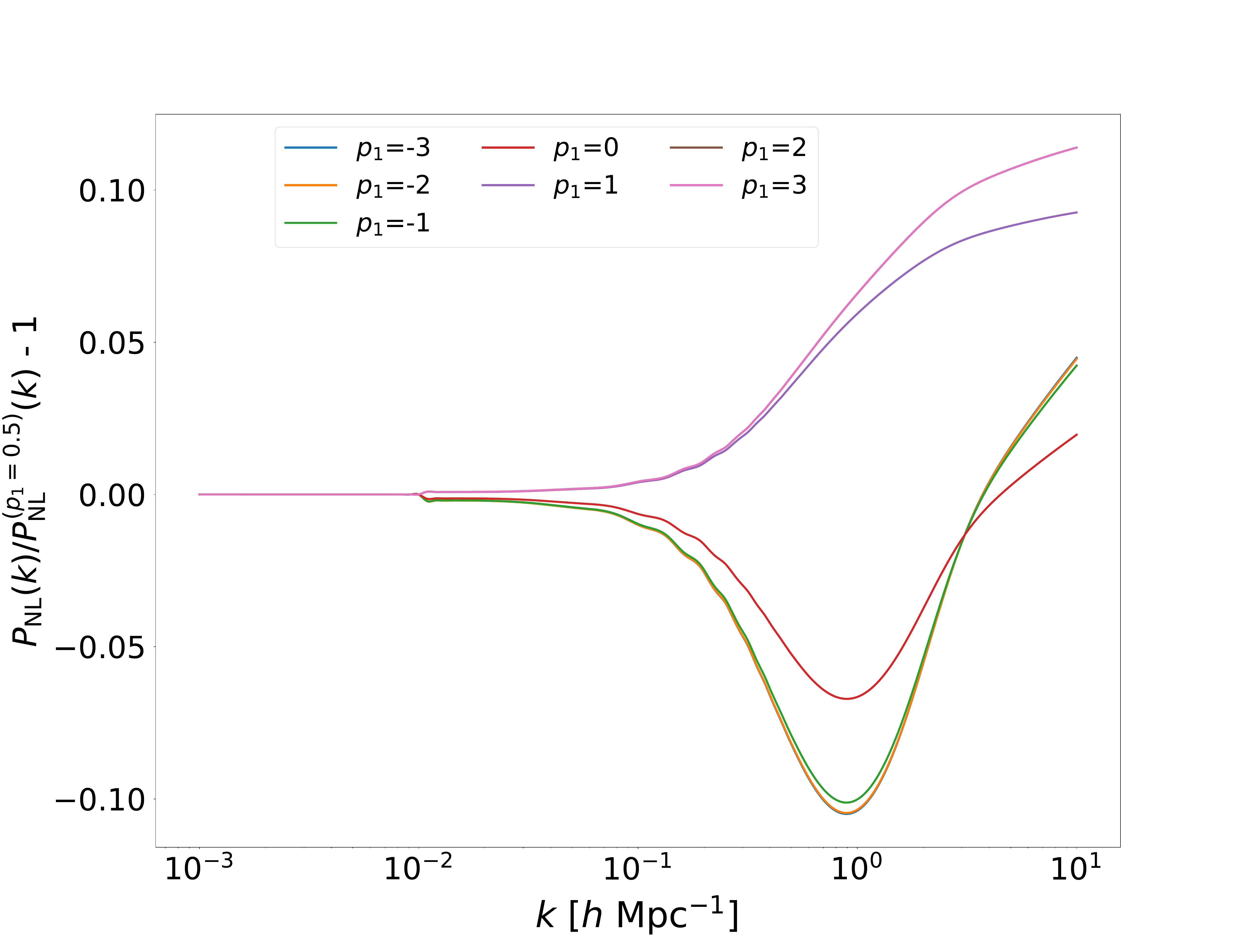}
    \caption{The effect of varying the reaction parameter $p_1$ on the nonlinear matter power spectrum $P_{\mathrm{NL}}(k)$. As reference, we use $p_1 = 0.5$ with the DES parametrization and $\mu_0=0.4$ and $\Sigma_0 = 0.1$.}
    \label{fig:test_NL_p1}
\end{figure}

We have observed that varying $p_1$ in the DES parameterization results in up to $\sim 10$\% difference in the nonlinear power spectrum at $z=0$, for $\mu_0=0.4$ and $\Sigma_0=0.1$, as shown by Fig.~\ref{fig:test_NL_p1}. The limit $p_1 \to \infty$ corresponds to no screening, whereas the limit $p_1 \to -\infty$ corresponds to strong screening, where the power spectrum goes back to the $\Lambda$CDM prediction.

\begin{figure}
    \centering
    \includegraphics[width = .5\textwidth]{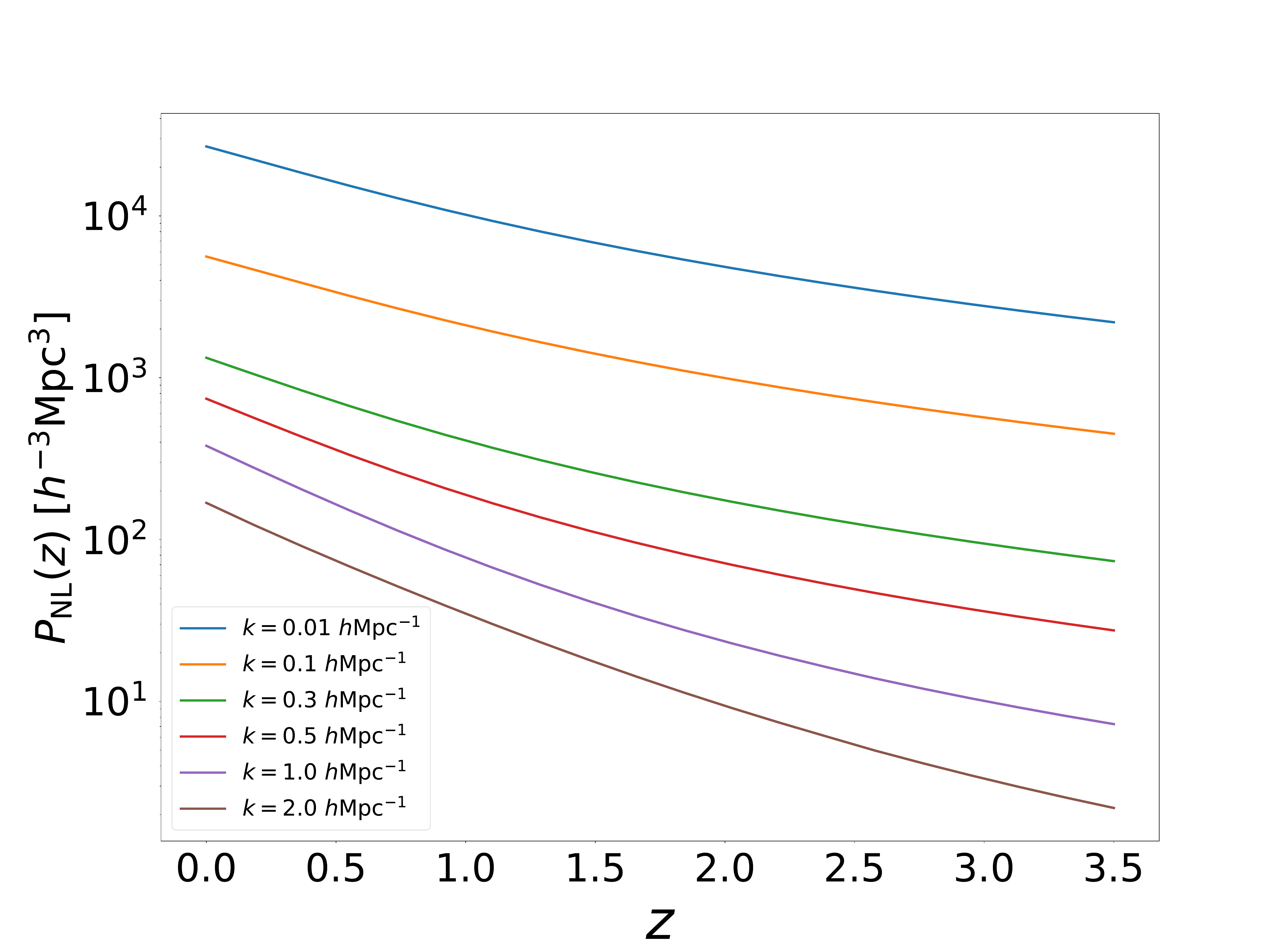}
    \caption{The modified matter power spectrum $P(k,z)$ as a function of redshift, for several representative values of $k$ and the DES parametrization, with $\mu_0=0.4, \Sigma_0=0.1$, and $p_1=0.5$. Note that we have a smooth transition around $z=2.5$, which is the redshift at which {\tt ReACT} corrections are activated.}
  \label{fig:test_NL_redshifts}
\end{figure}

The current version of {\tt ReACT} only makes predictions up to redshift $z=2.5$, which is sufficient to account for nonlinear modifications at late times. On the other hand, the redshift bins in the DES likelihood extend to $z=3.5$. To maintain a relatively smooth transition from low to high redshifts, we use $P^{\mathrm{pseudo}}(k,z)$ at $z>2.5$, without losing accuracy, given that most of the modified nonlinear corrections take place on redshifts below $z=2.5$. Fig.~\ref{fig:test_NL_redshifts} shows we do have a smooth transition in the output power spectrum from low to high redshifts, which we display for several representative $k$ modes.

\subsection{Other improvements} 
In addition to the nonlinear extension, we added a few other features to {\tt MGCAMB} and {\tt MGCobaya}, to aid cosmological tests of gravity.

\subsubsection{Galaxy-Weyl correlation in the DES Year-1 likelihood} \label{sec: feature-1}
The DES Year-1 likelihood, as implemented in {\tt Cobaya}, computes the galaxy-Weyl correlation from the galaxy-galaxy correlation, using the standard equations of GR relating $\delta$ and $W$.
In order to use the likelihood to constrain modifications of Einstein's equations, including the possibility of $\Sigma \neq 1$, we have modified the DES Year-1 likelihood implementation in {\tt MGCobaya} to compute $P_{W \mathrm{\delta}}(k)$  within {\tt MGCAMB}, together with $P_{W \mathrm{W}}(k)$ and $P_{\mathrm{\delta} \mathrm{\delta}}(k)$.

\subsubsection{Implementation of the DES Year-3 likelihood in \tt Cobaya}
We implement the DES Year-3 likelihood in {\tt Cobaya} by following the formulation for calculating the weak lensing observables presented in \cite{DES:2022ccp}. For the lens sample, we also follow the treatment in~\cite{DES:2022ccp}. Specifically, we choose the MagLim sample~\cite{Porredon:2021}, which contains six tomographic redshift slices with nominal edges at $z = [0.20, 0.40, 0.55, 0.70, 0.85, 0.95, 1.05]$, as the lens galaxy sample. We remove the two highest redshift MagLim bins from our analysis, as~\cite{Abbott:2022} revealed issues with the sample at $z > 0.85$.

Regarding the intrinsic alignments (IA) model used in the likelihood, we use the nonlinear
alignment (NLA) model,
which is consistent with the DES Collaboration's study of modified gravity constraints from their Year-3 data~\cite{DES:2022ccp}, which is not the same as the model (tidal alignment and tidal torquing, TATT) 
used in their constraints on the $\Lambda$CDM model~\cite{Abbott:2022}.

The two-point angular correlation functions for the separation $\theta$ are computed as:
\begin{equation}
\begin{aligned}\label{eq:xi_in_terms_of_Cell}
w^i(\theta) &= \sum_\ell \frac{2 \ell +1}{4\pi} P_\ell(\cos \theta) C^{ii}_{\delta_{\mathrm{obs}} \delta_{\mathrm{obs}}}(\ell) \, , \\
\gamma^{ij}_t(\theta) &= \sum_\ell \frac{2\ell + 1}{4\pi} \frac{P_\ell^2\left( \cos \theta \right)}{\ell(\ell + 1)} C_{\delta_{\mathrm{obs}}\mathrm{E}}^{ij}(\ell) \, , \\
\xi_{\pm}^{ij}(\theta) &= \sum_{\ell \geqslant 2} \frac{2\ell + 1}{4\pi}\ \frac{2(G_{\ell, 2}^+(\cos \theta) \pm G_{\ell, 2}^-(\cos \theta))}{\ell^2(\ell + 1)^2}\\ 
 & \times [ C^{ij}_{\mathrm{EE}}(\ell) \pm C^{ij}_{\mathrm{BB}}(\ell) ]\,.
\end{aligned}
\end{equation}
Here, $i,j$ denote two different redshift slices, $w^i$ is the two-point function between lens galaxy positions in redshift bin $i$, $\gamma^{ij}_t$ is the two-point function between lens galaxy positions and source galaxy tangential shear in redshift bins $i$ and $j$, and $\xi_{\pm}^{ij}$ is the correlation between source galaxy shears in redshift bins $i$ and $j$. $E$ and $B$ stand for the electric and magnetic parts of the shear field, respectively and $\delta_{\mathrm{obs}}$ is the observed projected galaxy density contrast. Furthermore, $P_\ell$ is the Legendre polynomials of order $\ell$, $P_\ell^m$ is the associated Legendre polynomial, and functions $G_{\ell,m}^{+/-}(x)$ are combinations of the associated Legendre polynomials $P_{\ell}^m(x)$ and $P_{\ell-1}^m(x)$, given explicitly in Eq.~(4.19) of~\cite{Stebbins:1996weak}.

The angular power spectra $C(\ell)$ in Eqs.~(\ref{eq:xi_in_terms_of_Cell}) receive contributions from the galaxy density perturbation ($\delta_{\mathrm{g}}$), gravitational shear ($\kappa$), intrinsic alignments (I), cosmic magnification (mag)
and redshift space distortions (RSD). Specifically,
\begin{align}\label{eq:cell}
\nonumber C^{ij}_{\mathrm{EE}}(\ell) = & C^{ij}_{\kappa\kappa}(\ell) + C^{ij}_{\kappa \mathrm{I_E}}(\ell) + C^{ji}_{\kappa \mathrm{I_E}}(\ell)+C^{ij}_{\mathrm{I_E}\mathrm{I_E}}(\ell) \, , \\
\nonumber C^{ij}_{\mathrm{BB}}(\ell) = & C^{ij}_{\mathrm{I_B}\mathrm{I_B}} \, , \\
\nonumber C^{ij}_{\delta_{\mathrm{obs}}\mathrm{E}}(\ell) = &C^{ij}_{\delta_{\mathrm{g}}\kappa}(\ell) + C^{ij}_{\delta_{\mathrm{g}}\mathrm{I_E}}(\ell) +  C^{ij}_{\delta_{\rm mag}\kappa}(\ell) + C^{ij}_{\delta_{\rm mag}\mathrm{I_E}}(\ell) \, , \\
\nonumber C^{ii}_{\delta_{\mathrm{obs}}\delta_{\mathrm{obs}}}(\ell) =&C^{ii}_{\delta_{\mathrm{g}}\delta_{\mathrm{g}}}(\ell)+C^{ii}_{\delta_{\rm mag}\delta_{\rm mag}}(\ell) +C^{ii}_{\delta_{\rm RSD}\delta_{\rm RSD}}(\ell) \\ &+ 2C^{ii}_{\delta_{\mathrm{g}}\delta_{\rm mag}}(\ell)+2C^{ii}_{\delta_{\rm g}\delta_{\rm RSD}}(\ell)+2C^{ii}_{\delta_{\rm RSD}\delta_{\rm mag}}(\ell)\,.
\end{align}
where $\mathrm{I_E}$ and $\mathrm{I_B}$ are the electric and magnetic parts of the intrinsic alignment, respectively.

The exact expression for the angular clustering power spectrum between two galaxy fields $A,B$ is
\begin{align}
\nonumber C_{\rm AB}^{ij} (\ell)=&\frac{2}{\pi}\int d \chi_1\,W^i_{A}(\chi_1)\int d\chi_2\,W^j_{B}(\chi_2)\\
    &\int\frac{dk}{k}k^3 P_{AB}(k,\chi_1,\chi_2)j_\ell(k\chi_1)j_\ell(k\chi_2)\,,
\label{eq:Cl-AB}
\end{align}
with $P_{\rm AB}$ being the corresponding three-dimensional power spectrum, and the kernels $W^{ij}_{\rm A,B}$ contain the relevant contributions mentioned above.

For the level of sensitivity of the DES measurements on the shear-shear ($C_{\mathrm{EE}}$, $C_{\mathrm{BB}}$) and galaxy-shear ($C_{\delta_{\mathrm{obs}}\mathrm{E}}$) spectra, one can evaluate them efficiently using the Limber approximation~\cite{Limber:1953}, namely, 
\begin{align}\label{eq:generalLimber}
C_{AB}^{ij}(\ell) = \int d\chi \frac{W_A^i(\chi)W_B^j(\chi)}{\chi^2}P_{AB}\left(k = \frac{\ell+\frac{1}{2}}{\chi},z(\chi)\right),
\end{align}
for which $P_{W\delta}$, $P_{W \mathrm{W}}(k)$ and $P_{\mathrm{\delta} \mathrm{\delta}}(k)$ are computed within {\tt MGCAMB}, as in the case of the DES Year-1 likelihood described in Sec.~\ref{sec: feature-1}.

However, the galaxy-galaxy spectrum ($C_{\delta_{\mathrm{g}}\delta_{\mathrm{g}}}$) is measured better, and the Limber approximation fails to provide the theory prediction with the required precision. Therefore we evaluate $C_{\delta_{\mathrm{g}}\delta_{\mathrm{g}}}$ using the exact formula in Eq. (\ref{eq:Cl-AB}). To be efficient, we follow the method described in Ref.~\cite{Fang:2020} to calculate the double-Bessel integral in Eq.~(\ref{eq:Cl-AB}) with the FFTLog algorithm. 

\subsubsection{Further improvements and synchronization with {\tt CAMB}}
We have created wrapper functions for $\mu$, $\gamma$, and $\Sigma$ in the Python interface of {\tt MGCAMB}, making it easy to test the time evolution of these phenomenological functions at a given Fourier number $k$ for all implemented modified gravity models.

The current version of {\tt MGCAMB}\footnote{\url{https://github.com/sfu-cosmo/MGCAMB}} is now forked with {\tt CAMB}\footnote{\url{https://github.com/cmbant/CAMB}} on GitHub, making it convenient to keep consistent with future upgrades of {\tt CAMB}.

\section{Demonstrations} \label{sec:demo}

To demonstrate the utility of the nonlinear extension of {\tt MGCAMB}, we consider the DES parametrization, to make it easier to compare against previous results in the literature. 

\begin{table*}[tbph]
 \centering
 \begin{tabular}{c|c|c}
  \hline
  \hline
  Parameter & Flat Prior & Fiducial \\ 
  \hline
  $\mu_0$ &  (-0.2,1.2) & 0.4 \\ 
  $\Sigma_0$ &  (-0.2,0.4)  & 0.1\\ 
  $p_1$ &  (-2,2) & 0.5 \\
  \hline
  $\Omega_b h^2$ &  - & 0.112 \\ 
  $\Omega_c h^2$ &  - & 0.0226 \\ 
  $n_s$ &  - & 0.969 \\ 
  $\mathrm{ln}(10^{10} A_s) $ &  - & 3.06 \\ 
  $100 \,\theta_{\mathrm{MC}}$ &  - & 1.0410 \\ 
  $\tau$ &  - & 0.067 \\ 
  $\Sigma\, m_{\nu}$ &  - & 0.06 \\ 
  \hline
  \hline
 \end{tabular}
\caption{Fiducial values and priors on the MG and cosmological parameters used in tests on synthetic DES-like data in Sec.~\ref{sec:demo}. Parameters other than $\mu_0,\Sigma_0$ and $p_1$ are fixed to their fiducial values, to allow us to isolate the impact of including the nonlinear extension to {\tt MGCAMB}.}
\label{tab: prior}
\end{table*}

\begin{table*}
 \centering
 \begin{tabular}{c|c|c|c}
  \hline
  & no NL ext + aggressive cut & NL ext + aggressive cut& NL ext + DES baseline \\
  \hline
  $\mu_0$&  0.392 $\pm$ 0.097 &  0.393 $\pm$ 0.098  &  0.393 $\pm$ 0.046  \\
  $\Sigma_0$  & 0.110 $\pm$ 0.015 & 0.102 $\pm$ 0.016  & 0.100 $\pm$ 0.012 \\
  \hline
 \end{tabular}
\caption{Tests on the nonlinear extension (NL ext). Parameter inferences on $\mu_0$ and $\Sigma_0$ obtained from synthetic data without the nonlinear extension and using the aggressive linear cut \textit{(first column)}, with the nonlinear extension while still using the aggressive linear cut \textit{(second column)}, and with the nonlinear extension and the DES baseline data, which includes nonlinear scales \textit{(third column)}. Results show that adding the nonlinear extension does not impact constraints derived from the linear scales alone, as expected (also see Fig.~\ref{fig:DES_mock_agg}). Note that constraints on $\mu_0$ are much tighter when nonlinear data is included (also see Fig.~\ref{fig:DES_mock_p1}).}
\label{tab:mock}
\end{table*}

First, we perform a test on mock data, generating synthetic DES-like galaxy-galaxy, galaxy-lensing and lensing-lensing (3$\times$2pt) correlations for a fiducial model with $\mu_0=0.4, \Sigma_0=0.1$ and $p_1=0.5$, and assuming a $\Lambda$CDM background expansion. The values of the other (cosmological) parameters are reported in Tab.~\ref{tab: prior}, whereas the DES likelihood parameters are fixed at the standard values of DES Year-1~\cite{DES:2018ufa, DES:2017tss}, to help us isolate the impact of including the nonlinear extension. As mentioned in Sec.~\ref{Sec:MGCAMB-with-ReACT}, the $p_1\rightarrow -\infty$ and $p_1 \rightarrow \infty$ limits of the nonlinear parameter $p_1$ describe the fully screened (i.e. the power spectrum goes back to the $\Lambda$CDM prediction) and fully unscreened regimes, respectively. In practice, we have verified that varying $p_1$ over the prior range $p_1\in[-2,2]$ is sufficient to capture this variability
, consistently with what shown in Ref.~\cite{Tsedrik:2024cdi}. This can be understood by noting that the transition from the screened to the unscreened regimes is modelled as an error function in the halo model reaction method \cite{Bose:2020wch}: at $p_1 \pm 2$, the error function has largely reached its asymptotes.

The mock data are generated using {\tt MGCAMB}. As our ``data'' covariance, we adopt the covariance matrix from the DES Year-1 likelihood, rescaled down by a factor of 25, as the actual DES Year-1 covariance would not allow us to constrain the nonlinear parameter $p_1$. We then run MCMC chains using {\tt MGCobaya}, to see if we can recover the input model. To test our implementation, and to assess the added constraints brought in by the nonlinear scales, we run with and without the nonlinear extension. In the latter case, we only vary the linear parameters $\mu_0$ and $\Sigma_0$, whereas we include $p_1$ when the nonlinear extension is applied. In the analyses we run, we employ the priors reported in Tab.~\ref{tab: prior}.

As mentioned earlier, the DES 3$\times$2pt data includes information from the nonlinear scales. In their Year-1 constraints on $\Lambda$CDM~\cite{DES:2018ufa}, the DES collaboration applied a conservative cut to keep only scales that could be reliably modelled with Halofit. We will refer to the DES 3$\times$2pt data with this conservative cut as ``DES baseline''. Conversely, in earlier {\tt MGCAMB} studies that used DES data, a more stringent cut was applied~\cite{Zucca:2019xhg} (denoted as ``aggressive cut'' in the following) that eliminated all the nonlinear scales -- as {\tt MGCAMB} was unable to model them prior to this work. Finally, when using DES Year-3 data, we will also consider a third type of cut: that used by the DES collaboration in the DES Year-3 constraints~\cite{DES:2022ccp} on modified gravity models, which also aims to restrict data to the linear scales. 

\begin{figure}[!htbp]
    \centering
    \includegraphics[width = .5\textwidth]{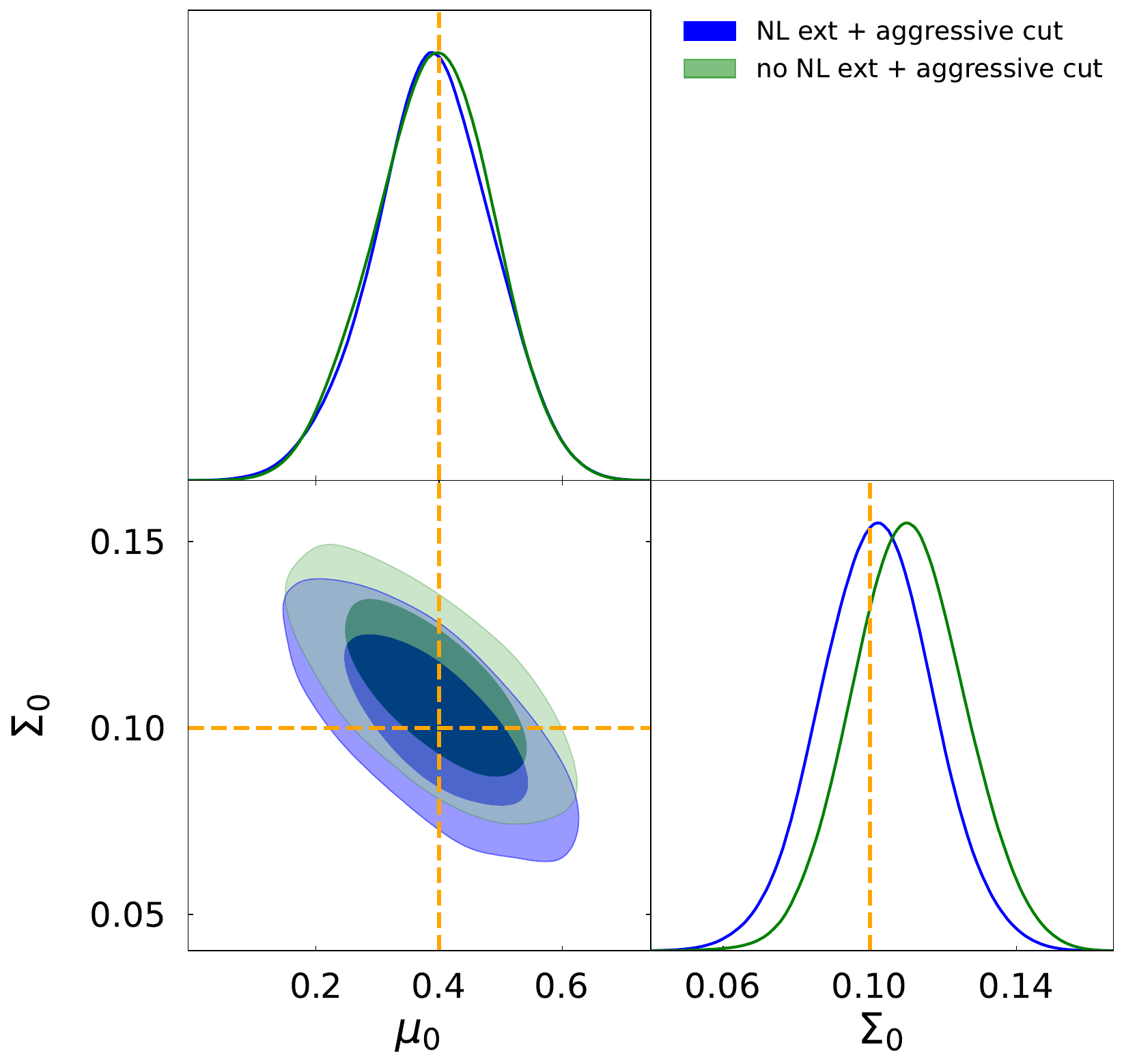}
    \caption{Test on the {\tt MGCAMB} extension. In this plot, we show the joint 68\% and 95\% CL constraints on $\mu_0$ and $\Sigma_0$ obtained from synthetic data with and without the nonlinear extension (NL ext), while applying the aggressive linear cut in both cases. Results show that the nonlinear extension does not impact constraints derived from linear scales alone, as expected.} 
    \label{fig:DES_mock_agg}
\end{figure}

To test the nonlinear extension to {\tt MGCAMB}, we first check that it does not impact constraints derived from the linear scales alone. To do so, we analyse the DES-like data with and without the nonlinear extension, while applying the aggressive cut. We show the results of this test in Fig.~\ref{fig:DES_mock_agg} and the first two columns of Tab.~\ref{tab:mock}: the recovered constraints on the parameters $\mu_0$ and $\Sigma_0$ are consistent with each other and with the fiducial model in both cases, demonstrating that indeed the nonlinear corrections does not impact the linear scales in unwanted ways. However, note that, when the nonlinear extension is disabled (green contours), the recovered values for the linear parameters $(\mu_0,\Sigma_0)$ are slightly shifted away from their true values: although small, this effect shows that nonzero values of the nonlinear parameter $p_1$ in the data might induce a bias in analyses that do not consider the effects of screening, even when limited to the linear scales.

\begin{figure}[!htbp]
    \centering
    \includegraphics[width = .5\textwidth]{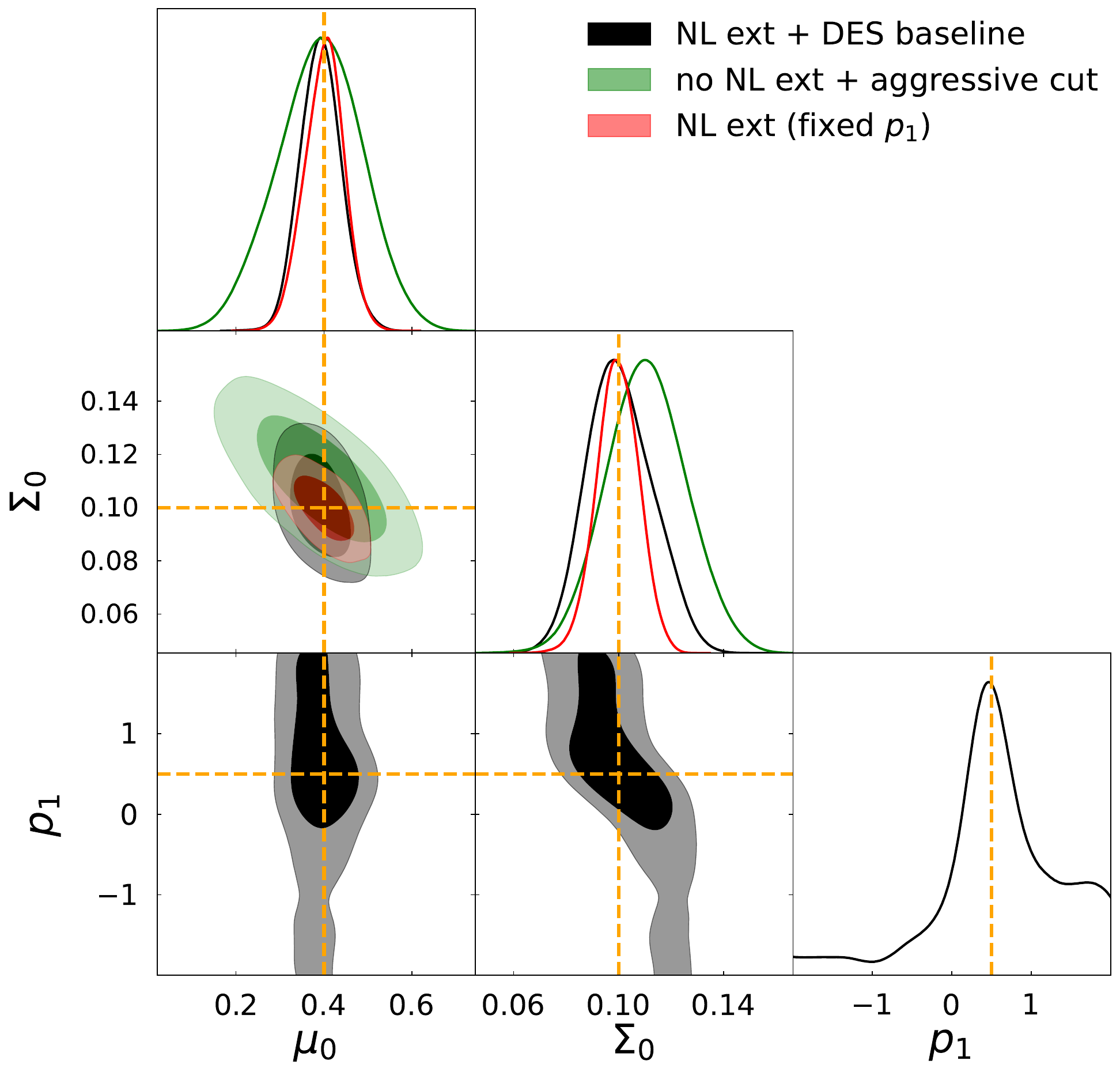}
    \caption{Comparison of the constraints on $\{\mu_0,\Sigma_0\}$ obtained with and without the nonlinear extension. In the latter case, the aggressive cut is applied. The constraints are derived from synthetic data generated from the fiducial parameters reported in Tab.~\ref{tab: prior}. Constraints are tighter when the nonlinear extension and the baseline data are used. We also show the constraint obtained by fixing $p_1$ to its fiducial value with the DES baseline synthetic data.} 
\label{fig:DES_mock_p1}
\end{figure}

As a second test, we assess the improvement in the constraints when we include the nonlinear scales. For this, we analyse the baseline DES-like synthetic data with and without the nonlinear extension. We also consider the case where $p_1$ is held fixed at the fiducial value. The joint constraints on $\mu_0$, $\Sigma_0$ and $p_1$ are shown in Fig.~\ref{fig:DES_mock_p1}. It is clear that we obtain better constraints on $\mu_0$ and $\Sigma_0$ when the nonlinear extension is used, even after marginalizing over $p_1$.
Since $p_1$ and $\mu_0$ do not degenerate strongly, the constraint on $\mu_0$ is only slightly improved by fixing $p_1$. On the other hand, the constraint on $\Sigma_0$ becomes tighter by fixing $p_1$ due to the breaking of the degeneracy between $\Sigma_0$ and $p_1$.
This demonstrates that the improvement is coming largely from the added scales in the ``DES baseline" dataset compared to ``aggressive cut". We also note that there is a bias in the recovered value of $\Sigma_0$ when the linear cut is applied. We report the parameter values we recover in Tab.~\ref{tab:mock}, showing that the $1\sigma$ uncertainty on $\mu_0$ is reduced by a factor of $\sim 2$ when the nonlinear extension is used, which agrees with Fig.3. of Ref.~\cite{Tsedrik:2024cdi}.

\begin{table*}[tbph]
 \centering
 \begin{tabular}{c|| >{\centering\arraybackslash}p{2cm}|| c c}
  \hline
  \multirow{2}{*}{\textbf{Parameter}} & \multirow{2}{*}{\textbf{Flat Prior}} & \multicolumn{2}{c}{\textbf{Posterior}} \\
    &  & DES Y3 baseline + PBRS & DES Y3 with linear cut + PBRS \\
  \hline
  $\mu_0$ &  (-1.0, 1.0)  & 0.003 $\pm$ 0.201 & -0.012 $\pm$ 0.216\\
  $\Sigma_0$ &  (-0.5, 0.5) & 0.016 $\pm$ 0.039 & 0.043 $\pm$ 0.045 \\
  $p_1$ &  (-2, 2)  &  -0.0006 $\pm$ 1.1283 & -  \\
  \hline
  $\Omega_b h^2$ &  (0.005, 0.1)  & 0.02245$\pm$ 0.00014 & 0.02247$\pm$ 0.00014 \\
  $\Omega_c h^2$ &  (0.001, 0.99)  & 0.1190$\pm$ 0.0009 &  0.1187$\pm$ 0.0009 \\
  $n_s$ &  (0.8, 1.2)  & 0.9692$\pm$ 0.0036 & 0.9690$\pm$ 0.0037 \\
  $\mathrm{ln}(10^{10} A_s) $ &  (1.61, 3.91) & 3.048$\pm$ 0.015 & 3.042$\pm$ 0.015 \\ 
  $100 \,\theta_{\mathrm{MC}}$ &  (0.5, 10)  & 1.0410$\pm$ 0.0003&  1.0411$\pm$ 0.0003\\ 
  $\tau$ &  (0.01, 0.8)  & 0.057$\pm$ 0.008 & 0.055$\pm$ 0.008  \\  
  \hline
  \hline
 \end{tabular}
\caption{The priors and the mean parameter values with the $68\%$ CL uncertainties derived from the combination of Planck 2018, BAO, RSD, and SN Ia (PBRS) data with DES Year-3 baseline data, with and without the nonlinear extension (also see Fig.~\ref{fig:DES_real_lin}). When excluding the nonlinear extension, we use the same linear cut as the DES collaboration \cite{DES:2022ccp}. Note the improvement on the constraints on the MG parameters when the nonlinear extension is included.}
\label{tab:real}
\end{table*}

\begin{figure}[!htbp]
    \centering
    \includegraphics[width = .5\textwidth]{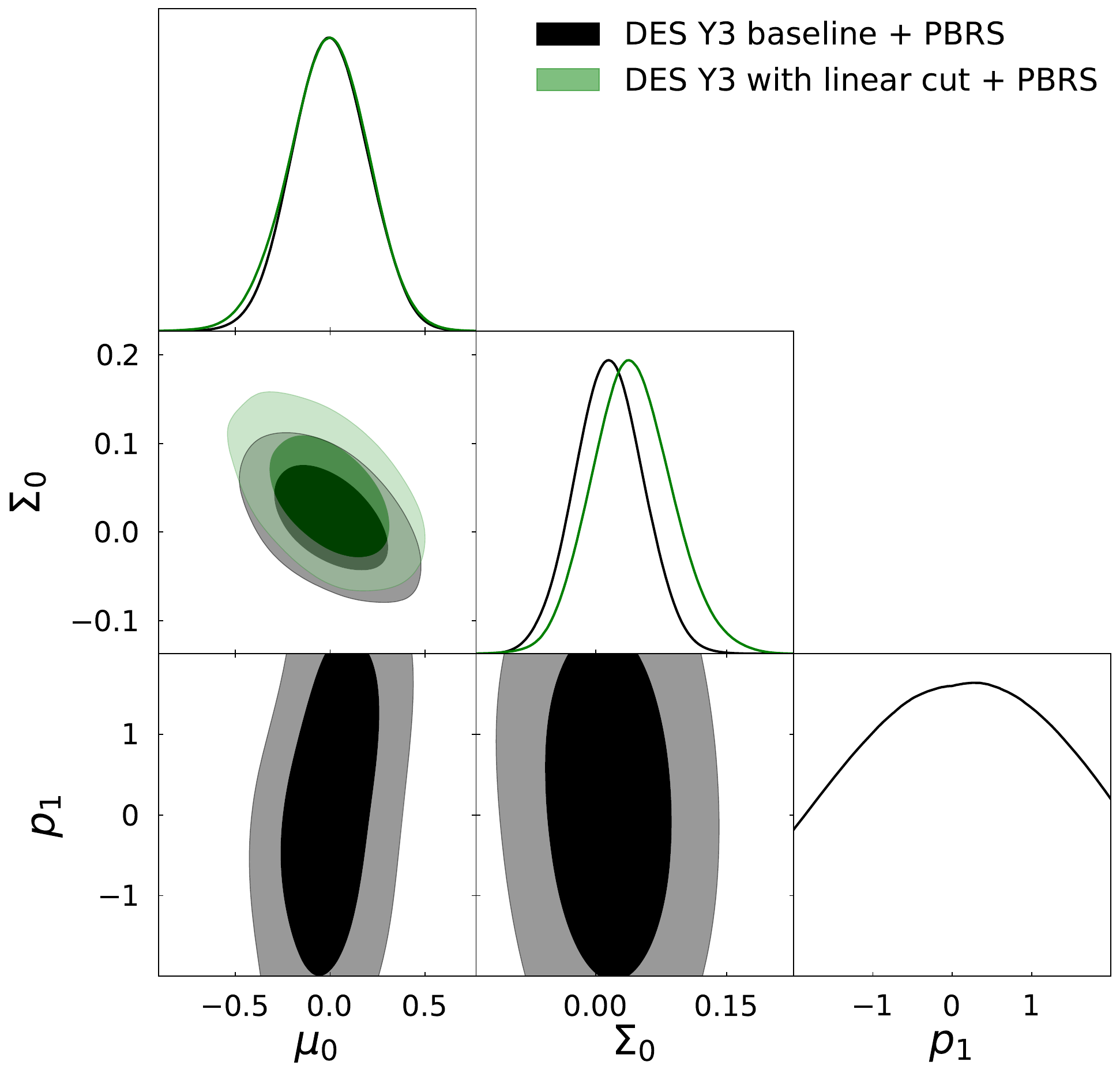}
    \caption{Constraints on $\mu_0$ and $\Sigma_0$ obtained from the baseline DES Year-3 data, with and without the nonlinear extension. In the latter case, we apply the same linear cut as the DES collaboration \cite{DES:2022ccp}. PBRS denotes the combination of Planck 2018, BAO, RSD, and SN Ia as detailed in Sec.~\ref{sec:demo}. Constraints on $\Sigma_0$ improve when the nonlinear extension is used.}
    \label{fig:DES_real_lin}
\end{figure}

Having tested our setup, we apply the extended  {\tt MGCAMB} to current cosmological datasets, deriving constraints on $\mu_0$ and $\Sigma_0$, along with the main cosmological parameters. For this, we use the DES Year-3 3$\times$2pt data~\cite{DES:2022ccp}, Planck 2018 CMB temperature, polarization and lensing power spectra~\cite{Aghanim:2019ame}, joint measurements of baryon acoustic oscillations (BAO) and redshift-space distortions (RSD) from eBOSS DR16, the SDSS DR7 MGS data~\cite{Ross:2014qpa}, the BAO measurement from 6dF~\cite{Beutler_2011}, and the Pantheon sample of uncalibrated supernovae~\cite{Pan-STARRS1:2017jku}. In addition to $\mu_0$ and $\Sigma_0$, we vary all the standard cosmological parameters: the baryon density $\Omega_b h^2$, the CDM density $\Omega_c h^2$, the amplitude $A_s$ and the spectral index $n_s$ of the primordial fluctuations spectrum, the angular size of the sound horizon at decoupling $\theta_{\mathrm{MC}}$ and the optical depth $\tau$. To assess, the impact of including nonlinear scales, we perform two analyses, with and without the nonlinear extension, applying the DES Year-3 linear cut in the latter. We show the recovered posteriors for $\mu_0$, $\Sigma_0$ and $p_1$ in Fig.~\ref{fig:DES_real_lin}: it is clear that we obtain stronger constraints on $\mu_0$ and $\Sigma_0$ with the nonlinear extension. In fact, whilst the posterior on $\mu_0$ is largely unaffected, the improvement on the $\Sigma_0$ constraint is apparent. We can understand this by noting that DES is primarily a weak lensing survey, which primarily constraints $\Sigma$. The priors and the mean values with uncertainties for all parameters are provided in Tab.~\ref{tab:real}. The constraints we obtain on $\mu_0$ and $\Sigma_0$ with the DES Year-3 linear cut are consistent with those reported by the DES collaboration \cite{DES:2022ccp}.

Finally we make a note on baryonic effects, known to play a large role in the nonlinear regime. These effects have been found to be largely degenerate with the modified-gravity screening parameter $p_1$ in Ref.~\cite{Tsedrik:2024cdi}, where the authors provided forecasts much deeper into the nonlinear regime than considered here. As this work is meant to be a demonstration of the new implementation in {\tt MGCAMB}, we do not consider very nonlinear scales and we do not include baryonic effects. We note that the baseline scale cut developed by the DES collaboration was chosen to avoid contamination from baryonic effects. This can be confirmed by the fact that $p_1$ is not well constrained by the DES Year-3 3x2pt data. However, with the covariance rescaled by a factor of 25, $p_1$ is well constrained, and this means that baryonic effects will play a significant role. It should  be noted that it may be sufficient to include a $\Lambda$CDM-derived baryonic boost factor along with the prediction of $P_{\rm NL}(k,z)$ even in MG scenarios, as supported by various recent studies \citep{Mead:2016zqy,Arnold:2019zup,Hernandez-Aguayo:2020kgq}, which show that baryonic and MG effects are largely decoupled.

\section{Summary} 
\label{sec:summary}

We have extended {\tt MGCAMB} to predict cosmological observables in modified gravity at the nonlinear scales. The extension employed the halo model reaction method, with the reaction function computed using {\tt ReACT} and interfaced in the Python wrapper of {\tt MGCAMB}. 

We have tested the nonlinear extension on synthetic data generated using the DES parameterization, a two-parameter phenomenological model, recovering the input fiducial parameters $\mu_0$ and $\Sigma_0$. Further tests showed that constraints on $\mu_0$ and $\Sigma_0$ improve when the new extension is included.


Having tested our method, we then used it to obtain constraints on $\mu_0$ and $\Sigma_0$ from a combination of current datasets that included the DES Year-3 data, and found an enhancement in the constraints compared to results obtained from the linear scales only.

This nonlinear extension of {\tt MGCAMB} will enable us to use a wider range of data from current and forthcoming cosmological surveys, including weak lensing surveys such as the Vera C. Rubin Observatory\footnote{\url{https://www.lsst.org/}} and Euclid\footnote{\url{https://www.esa.int/Science_Exploration/Space_Science/Euclid}}. 

\vspace{2cm}
\acknowledgments 
The work of ZW and LP is supported by the National Sciences and Engineering Research Council (NSERC) of Canada. LY and GBZ are supported by the National Key Basic Research and Development Program of China (No. 2018YFA0404503), NSFC (11925303, 11890691), and science research grants from the China Manned Space Project with No. CMS-CSST-2021-B01. DS and KK are supported by STFC grant ST/W001225/1. BB is supported by a UKRI Stephen Hawking Fellowship (EP/W005654/2). 
For the purpose of open access, we have applied a Creative Commons Attribution (CC BY) licence to any Author Accepted Manuscript version arising. Supporting research data are available on reasonable request from the corresponding author.
\vspace{2cm}


%

\end{document}